\documentclass[12pt,a4paper]{article}
\usepackage[english]{babel}
\usepackage{indentfirst}
\usepackage{graphicx}
\usepackage{subfigure}

\begin{document}

\begin{center}
{\bf {ELLIPTICAL MOTIONS OF STARS IN CLOSE BINARY SYSTEMS}}
\end{center}

\begin{center}
\copyright \ { L.\,G.\,Lukyanov}\footnote{luka@sai.msu.ru}, S.\,A.\,
Gasanov \footnote{gasanov@sai.msu.ru}\\
{\small{\it Sternberg State Astronomical Institute, Moscow}}
\end{center}

\begin {center}
Abstract.\end {center}

Motions of stars in close binary systems with a conservative mass
exchange are examined. It is shown that Paczynski-Huang model widely
used now for obtaining the semi-major axis variation of a relative
stars orbit is incorrect, because it brings about large mistakes. A
new model suitable for elliptical orbits of stars is proposed. Both
of reactive and attractive forces between stars and a substance of
the flowing jet are taken into account. A possibility of a mass
exchange at presence of accretion disk is considered.


\begin{center} INTRODUCTION \end{center}

The research of stars motions in close binary systems began in
the 60-s' of the last century in a cycle of works by Kruszevski
[1], Hadjidemetriou [2], Piotrovski [3], Huang [4], Paczynski [5]
and others. As a result for a case of conservative mass exchange the simplified dependence was received for semi-major axis $a$ of a circular relative stars orbit from constant mass increase of an accepting star
$\dot {M} _ 2$ as
\begin {equation} \label {V1}
\dot {a} = 2a\dot {M} _ 2\left (\frac {1} {M _ 1} -\frac {1} {M _ 2}
\right).
\end {equation}

This dependence, which we shall call for by Paczynski-Huang name and up
to this day it is used in all researches dedicated to close binary systems with a
conservative mass exchange when
$M_1 + M_2 = M = \mathrm
{const} $.

The derivation of formula (1) is based on the assumption that
equations of motion for stars with variable masses admit
the angular momentum integral
\begin {equation} \label {V2}
\mathbf {J} = M _ 1\mathbf {R} _ 1\times \mathbf {V} _ 1 + M _
2\mathbf {R} _ 2\times \mathbf {V} _ 2 = \mathbf {const},
\end {equation}
where $M_i, \mathbf {R}_i, \mathbf {V}_i, \; (i = 1, 2) $ --- the
mass, radius-vector and velocity of stars motion. By the most an
assumption is made that the close binary stars form a closed
mechanical system which admits integrals of momentum and angular
momentum. Differential equations of motion thus are usually not
written out.

However the assumption on existence of integral of momentum is
erroneous. We shall be convinced of it, proceeding from most general
view of the Mestschersky equations for a two-body problem with
variable masses [6]:
\begin {equation} \label {a}
M _ 1\frac {d {\bf V}_1} {d \, t} = G \frac {M_1 \, M_2} {R^
3} {\bf R} + {\bf Q}_1, \qquad M _ 2\frac {d {\bf V}_2} {d \, t}
= - G \frac {M_1 \, M_2} {R^3} {\bf R} + {\bf Q} _ 2,
\end {equation}
where $ \bf R = \bf R _ 2 - \bf R _ 1 $, and $ {\bf Q} _ 1 $ and $
{\bf Q} _ 2 $ are jet forces acting on stars $S_1$ and $S_2$.

Let's consider three opportunities of reception of integral of
momentum (\ref {V2}) from equations (\ref {a}).

1. At first we shall take into account of closed mechanical system,
described by the problem of Mestschersky-Levy-Civita [7], differential equations of which turn out from (\ref {a}), if jet forces put
equal to:
\begin {equation} \label {b}
{ \bf Q} _ 1 = -\dot {M} _ 1 {\bf V} _ 1, \qquad {\bf Q} _ 2 = -\dot
{M} _ 2 {\bf V} _ 2,
\end {equation}
where $ \dot {M} _ 1 = -\dot {M} _ 2 = \mathrm {const} $, $ \dot {M}
_ 2> 0 $.

This model is usually used in astronomy at study of motions of
mutually attracted pair stars, taking place in dust cloud with
account of jet forces, arising owing to sticking substances dust of
cloud on stars.

The Mestschersky-Levy-Civita problem represents a closed mechanical
system, therefore there are integrals of momentum and of angular
momentum (\ref {V2}). However to use model Mestschersky-Levy-Civita
for study on motions of stars in close binary system with a
conservative exchange of mass is not admitted, so as the jet forces
(\ref {b}) in close binary system do not exist. The true jet forces
have other directions and others absolute values: on a donor-star
the jet force acts, directed in the opposite side to radius $ {\bf
R} $ and is equal to product of velocity of sound on $\dot{M}_1$,
and on accreted-star acts the jet force, directed on tangent to a
trajectory of relative motion of flowing particles at the moment of
their hit on a surface of a star. These jet forces sharply differ
from values (\ref {b}) in the Mestschersky-Levy-Civita problem.

For this reason the model of Mestschersky-Levy-Civita is not suitable
for study of stars motion in close binary system.

2. Any other problems of two bodies with variable masses, admitting
integral of angular momentum, does not exist. But it
is possible to consider an opportunity of the approximate reception
of integral (\ref {V2}) from equations (\ref {a}). For this
purpose we shall consider, that the jet forces acting on stars, are
small values in comparison with forces of mutual attraction between
stars, and consequently they can be neglected
\begin {equation} \label {c}
{ \bf Q} _ 1 = 0, \qquad {\bf Q} _ 2 = 0.
\end {equation}

Then we shall receive a well known Gylden-Mestschersky model differential equations of which are
\begin {equation} \label {}
M _ 1\frac {d {\bf V} _ 1} {d \, t} = G \frac {M _ 1 \, M _ 2} {R ^
3} {\bf R}, \qquad M _ 2\frac {d {\bf V} _ 2} {d \, t} = - G \frac
{M _ 1 \, M _ 2} {R ^ 3} {\bf R},
\end {equation}
in which integral of the angular momentum (\ref {V2}) in
absolute motion does not exist, but a quasi-integral does exist
\begin {equation} \label {d}
\mathbf {J} ' = M _ 1\mathbf {R} _ 1\times \mathbf {V} _ 1 + M _
2\mathbf {R} _ 2\times \mathbf {V} _ 2 -\int (\dot {M} _ 1\mathbf
{R} _ 1\times \mathbf {V} _ 1 + \dot {M} _ 2\mathbf {R} _ 2\times
\mathbf {V} _ 2) \, dt = \mathbf {const}.
\end {equation}

If in this quasi-integral one neglect small members, containing as multiplier the velocity of mass change $ \dot {M} _ 1 $ and $
\dot {M} _ 2 $, we shall receive integral (\ref {V2}), but thus a
problem of Gylden-Mestschersky will be transformed to a two-body
problem with constant masses, with the help of which it is impossible to
study a transfer of substance between stars.

Nevertheless  the conservative ($ M = \mathrm {const} $) problem of
Gylden-Mestschersky represents the certain interest for study
conservative exchange in close binary system, so as it is possible
with the help of it to characterize approximately motions of stars
at presence of accretion disk. The equations of relative motion in
the Gylden-Mestschersky problem are
\begin {equation} \label {}
\frac {d\mathbf {V}} {dt} = -\frac {GM} {R ^ 3} \mathbf {R}
\end {equation}
and at $ M = \mathrm {const} $ exactly coincide with the equations of
relative motions of the classical two-body problem with constant
masses [8], therefore for relative motion of bodies in the
problem of Gylden-Mestschersky there is a strict angular momentum
integral
\begin {equation} \label {}
\sqrt {GMa (1-e ^ 2)} = \mathrm {const},
\end {equation}
and all Kepler's elements of an orbit in this problem are constants,
including $ a = \mathrm {const} $.

Thus, the exclusion from consideration all of jet forces, i.e. the
transition to a conservative problem of Gylden-Mestschersky, results
to invariance of the semi-major axis of a relative orbit of stars,
but does not result in the formula of Paczynski-Huang.

3. We shall consider now the third opportunity of a "formal
conclusion"\, the rules (\ref {V1}), which is usually used in the
literature. Proceeding from the lack of external forces acting on a
system, and invariance of its complete mass, one asserts, that the
system is closed and, hence, admits the existence of integral (\ref
{V2}).

Such statement is valid for stars with constant masses, but is not valid for
stars with variable masses. The point is that for systems of
bodies with variable masses to be closed except of absence of external forces
and the invariance of complete mass is needed it is  necessary a presence some additional "internal"\, jet forces, which make system  to be closed.

Really, according to the second law of Newtons mechanics the velocity
speeds of change of momentum of stars are determined by
equalities:
\begin {equation} \label {}
\frac {d (M _ 1\mathbf {V} _ 1)} {dt} = G \frac {M _ 1 \, M _ 2} {R
^ 3} {\bf R} + \mathbf {F} _ 1, \quad \frac {d (M _ 2\mathbf {V} _
2)} {dt} = - G \frac {M _ 1 \, M _ 2} {R ^ 3} {\bf R} + \mathbf {F}
_ 2,
\end {equation}
where $ \mathbf {F} _ 1 $ and $ \mathbf {F} _ 2 $ are external
forces, acting on stars.

Let's rewrite these equations as
\begin {equation} \label {}
M _ 1\frac {d\mathbf {V} _ 1} {dt} = G \frac {M _ 1 \, M _ 2} {R ^
3} {\bf R} + \mathbf {Q} _ 1 + \mathbf {F} _ 1, \quad M _ 2\frac
{d\mathbf {V} _ 2} {dt} = - G \frac {M _ 1 \, M _ 2} {R ^ 3} {\bf R}
+ \mathbf {Q} _ 2 + \mathbf {F} _ 2,
\end {equation}
where $ \mathbf {Q} _ 1 = -\dot {M} _ 1 \mathbf {V} _ 1 $ and $
\mathbf {Q} _ 2 = -\dot {M} _ 2 \mathbf {V} _ 2 $ are "internal"\,
jet forces, active on stars $ S _ 1 $ and $ S _ 2 $.

From there it is visible, that integral of momentum
$M_1\mathbf {V} _ 1 +  M _ 2\mathbf {V} _ 2 = \mathbf {const} $ and
together with it those of angular moment (\ref {V2}), will
exist at absence of external forces, $ \mathbf {F} _ 1 = \mathbf {F} _
2 = 0 $, but it is obviously necessary a preservation of "internal"\, jet
forces $\mathbf{Q}_1=-\dot{M}_1 \mathbf{V}_1$ and $\mathbf{Q}_2 =
-\dot{M}_2 \mathbf{V}_2$.

Therefore for existence of integral (\ref {V2}) it is necessary not
only the absence of external forces and invariance of complete mass
of the system, but also the presence of "internal"\, jet forces of a
kind (\ref {b}) too is necessary. These requirements are carried out
only for the problem of Mestschersky-Levy-Civita, considered in item
1. For a problem on motion of stars in close binary system with a
conservative exchange of mass these requirements are not carried
out, so as the jet forces have another vector values which are
distinct from "internal"\, jet forces, namely, $ \mathbf {Q} _ 1 =
-\dot {M} _ 1\mathbf {W} _ 1 $ and $ \mathbf {Q} _ 2 = -\dot {M} _
2\mathbf {W} _ 2 $, where $ \mathbf {W} _ 1\neq \mathbf {V} _ 1 $
and $ \mathbf {W} _ 2\neq \mathbf {V} _ 2 $ are relative velocities
of the outflow (inflow) of mass on stars. Therefore the integral of
angular momentum does not here exist.

Thus, the assumption on existence of integral of angular momentum
(\ref {V2}) is infaithfull, i.e. those about a problem on stars
motion in close binary system with a conservative exchange of mass
being closed, is erroneous. In a consequence of that the model of
Paczynski-Huang is incorrect, and it cannot be used for study of
motions in close binary stars. As show results of numerical
integration of equations of motion with taking into account of true
jet forces, acting on a star, the use of Paczynski-Huang model
results in significant mistakes in definition of the semi-major
axis, down to an opposite sign of derivative $ \dot {a} $.

For this reason any conventional correct model, determining
motions of close binary stars, now does not exist. By the aim of purpose our
work is the such model creation. For circular motions of stars such
a model was offered in the work [9], in which except for forces of mutual
attraction of stars both the true jet forces, and forces of attraction on stars from jet stream of substance are taken into account
also. The definition of stars motions is carried out with the help of
numerical integration. In the present work a similar research, but under assumption, that the orbits of stars are elliptic will be carried out.

\begin {center} THE RESTRICTED ELLIPTICAL THREE-BODY PROBLEM \end {center}

For definition of motion of flowing particles of masses we shall use the restricted elliptical three-body problem, the equations of which in rotating and pulsating, barycentric system of coordinates of Schepner (Scheibner, Petr, Nechvil, Rein) $x, y$ look
like [10]
\begin {equation} \label {1.1}
\begin {array} {c}
\displaystyle\frac {d ^ 2x} {dv ^ 2} -2\frac {dy} {dv} = \rho\frac
{\partial U} {\partial x}, \\ [5pt] \displaystyle\frac {d ^ 2y} {dv
^ 2} + 2\frac {dx} {dv} = \rho\frac {\partial
U} {\partial y}, \\
\end {array}
\end {equation}
where the $x$-axis is always directed on a accreted-star $ S _ 2 $,
$v$ is true anomaly of stars, $\rho = 1/(1 + e\cos v)$ is the
dimensionless distance of a star $S_2$ comparatively $S_1$. Through
$ U $ the Jacobi function in the restricted elliptic three-body
problem is designated
\begin {equation} \label {1.2}
 U = \frac {x ^ 2 + y ^ 2} {2} + p ^ 3\left (\frac {1-m} {r _ 1} + \frac {m} {r _ 2} \right)
+ \frac {p ^ 2} {2}(3 + m ^ 2-m),
\end {equation}
and $r_1$ and $r_2$ are distances from a flowing particle of a jet
accordingly up to the centers of masses of the first and second
stars
\begin {equation} \label {1.3}
r_1 = \sqrt{(x + pm)^2 + y^2}, \quad r_2 = \sqrt{(x + pm - p)^2 +
y^2},
\end {equation}
where $1-m = M_1/M$ and $m = M_2/M$ are relative masses of stars, $M
= M_1 +M_2$, $p = a(1-e^2)$ is the focal parameter of an orbit, $a$
and $e$ are semi-major axis and eccentricity accordingly.

The stream of substance from a donor-star occurs through a vicinity
of internal Euler libration point $L_1$, àáñöèññà $x_L$ with which
is conclude in limits $ -pm < x _ L < p-pm $ and is determined
numerically as a root of the nonlinear equation $
\partial U/\partial x = 0 $ at $ y = 0 $, i.e. of equation
\begin {equation} \label {}
x - \displaystyle p ^ 3\left (\frac {1 - m} {\sqrt {x + pm}} +
\displaystyle \frac {m} {\sqrt {x + pm - p}} \right) = 0.
\end {equation}

Roche lobes in the planar elliptic restricted three-body problem are
determined with the help of equations of curves of minimal energy
[11]
\begin {equation} \label {}
x^2 + y ^ 2 + 2p ^ 3\left (\frac {1-m} {r _ 1} + \frac {m} {r _ 2}
\right) -p ^ 2 (3 + m ^ 2-m) = C (1 + e\cos v),
\end {equation}
where $C$ is Jacobi's constant.

Let's consider, that the accepting star has the form of sphere
\begin {equation} \label {1.7}
\rho ^ 2[(x + pm-p) ^ 2 + y ^ 2] = P ^ 2,
\end {equation}
the radius $P$ of which in stream process of substance changes
according to the dependence
\begin {equation} \label {1.8}
P = P _ 0 \sqrt [3 \;] {\frac {m} {m _ 0}}.
\end {equation}

The stream of substance begins with a donor-star after achievement
by flowing particles a level of energy more than that of in the
lipration point $L_1$. As it has been shown in [11], the stream of
substance through the vicinity of a point $L_1$ has a pulsating
character and occurs in the vicinity of apoastr of orbits. The
velocity $V_0$ of the stream of particles masses from a star $S_1$
is always directed along the $x$-axis to the star $S_2$ and has the
value
\begin {equation} \label {1.5}
V_0(v) = V_{00} V_1(v), \quad \displaystyle V_1(v) = m
\sqrt{\frac{GM}{p}}\sqrt{1+2e\cos v +e^2},
\end {equation}
where $ V _ {00} \simeq 0.03 $ is the coefficient, established from
observations. The apoastr vicinity, in which occurs the stream of
mass, is determined on a level of energy of particles (Jacobi's
constant $C$, see [11]) by range of true anomaly $\pi - v_a \leq v
\leq \pi + v_a$, where $ 0 < v _ a < \pi $. Therefore the mean
velocity $V_c$ of stream of mass from the star $S_1$ during one
orbital period of stars is possible to determine with the help of
formula
\begin {equation} \label {1.5a}
V _ c = \frac {1} {2v _ a} \int\limits _ {\pi-v _ a} ^ {\pi + v _ a}
V _ 0 \, dv = V _ {00} m \sqrt {\frac {GM} {p}} \frac {1} {2v _ a}
\int\limits _ {\pi-v _ a} ^ {\pi + v _ a} \sqrt {1 + 2e\cos v + e ^
2} \, dv.
\end {equation}

All stream lines of a jet, having scatter both on coordinates and on
time during one "emission"\, of substance for an orbital period, we
shall approximate by one trajectory, outgoing from singular point
$L_1$. The numerical integration of equations (\ref {1.1}) is
carried out with such initial conditions:
\begin {equation} \label {1.8}
\begin {array} {c}
\displaystyle v _ 0 = \pi, \; v _ a = \frac {\pi} {2}, \; x (v _ 0)
= x _ L, \; x ' (v _ 0) = \frac {V _ c} {\dot {v}},
\; y (v _ 0) = 0, \; y ' (v _ 0) = 0, \\
m = m_0, \; \dot {m} = \mathrm {const}, \; P = P _ 0, \\
\end {array}
\end {equation}
where "prime"\, signifies differentiation on true anomaly, $ (\dots)
' = d (\dots) /dv $.

The numerical integration is carried out on an interval $ v _ 0\leq
v\leq v _ 0 + \tau $, where the value of true anomaly $ v _ 0 + \tau
$ is determined by the moment of hit of a particle on the second
star surface in the point $x_2 = x (v _ 0 + \tau), y _ 2 = y (v _ 0
+ \tau) $. If such a moment does not exist, the process of stream of
substance occurs with forming of an accretion disk.

As a result of numerical integration there are components of outflow
velocity of $\mathbf{W}_1$ from the star $S_1$ and the inflow mass
velocity $\mathbf{W}_2$ on the star $S_2$:
\begin {equation} \label {1.9}
W _ {1x} = V _ c, \; W _ {1y} = 0, \; W _ {2x} = x ' (v _ 0 + \tau)
\dot {v}, \; W _ {2y} = y ' (v _ 0 + \tau) \dot {v}.
\end {equation}

Jet forces acting on stars $S_1$ and $S_2$ are considered to be
applicable to their centers of masses and are determined with these
formulas
\begin {equation} \label {1.10}
\mathbf {Q} _ 1 = \dot {m} M \{-V _ c, 0 \}, \quad \mathbf {Q} _ 2 =
\dot {m} M \{x ' (v _ 0 + \tau) \dot {v}, y ' (v _ 0 + \tau) \dot
{v} \}.
\end {equation}

With the help of numerical integration also the mass of established
stream $S_3$ and coordinates $x_3, y_3$ of its center of masses are
determined as:
\begin {equation} \label {1.11}
M _ 3 = \frac {\tau} {\dot v _ c} \dot m M, \quad X_3 = \frac {1}
{\tau} \int\limits _ {v _ 0} ^ {v _ 0 + \tau} x(v) dv, \quad Y _ 3 =
\frac {1} {\tau} \int\limits _ {v _ 0} ^ {v _ 0 + \tau} y(v) dv,
\end {equation}
where
$$
\dot v _ c = \frac {1} {2\pi} \int\limits _ 0 ^ {2\pi} \dot v \, dv
= \frac {1} {2\pi} \int\limits _ 0 ^ {2\pi} \sqrt {\frac {GM} {p ^
3}} \, \left (1 + e \cos v \,\right) ^ 2 \, dv = \displaystyle \sqrt
{\frac {GM} {p ^ 3}} \, \left (1 + \frac {e ^ 2} {2} \, \right)
$$
is the mean angular velocity of orbital motion.

With the help of formulas (\ref {1.10}) and (\ref {1.11}) the jet
forces, the mass of jet and coordinates of its center of masses are
determined. These values enter in the right-hand sides of the
differential equations of relative motion of stars.

\begin {center}
DIFFERENTIAL EQUATIONS OF MOTION OF STARS
\end {center}

As initial differential equations of motion of stars in an inertial
coordinate system we shall consider system of equations:
\begin {equation} \label {2.1}
\begin {array} {c}
\displaystyle M _ 1\frac {d {\bf V} _ 1} {d \, t} = G \frac {M _ 1
\, M _ 2} {R ^ 3} {\bf R} + \mathbf {Q} _ 1 + G\frac {M _ 1M _ 3} {r
_ {13} ^ 3} \mathbf {r} _ {13}, \\ [10pt] \displaystyle M _ 2\frac
{d {\bf V} _ 2} {d \, t} = - G \frac {M _ 1 \, M _ 2} {R ^ 3} {\bf
R} + \mathbf {Q} _ 2 + G\frac {M _ 2M _ 3} {r _ {23} ^ 3}
\mathbf {r} _ {23}, \\
\end {array}
\end {equation}
where $r_{13} = \sqrt {(x _ 3 + pm) ^ 2 + y _ 3 ^ 2}, \quad r_{23} =
\sqrt {(x _ 3 + pm-p) ^ 2 + y _ 3 ^ 2} $ distances between centers
of masses of stars and that of the jet. The equation of a star $S_2$
relative $S_1$ motion from there turns out to be as
\begin {equation} \label {2.2}
\frac {d {\bf V}} {d \, t} = - \frac {GM} {R ^ 3} {\bf R} + \frac
{\mathbf {Q} _ 2} {M _ 2} -\frac {\mathbf {Q} _ 1} {M _ 1} + GM _
3\left (\frac {\mathbf {r} _ {23}} {r _ {23} ^ 3} -\frac {\mathbf
{r} _ {13}} { r _ {13} ^ 3} \right),
\end {equation}
where $\mathbf{V} = {\bf V}_2 - {\bf V}_1$ is the velocity of a star
$S_2$ relative to $S_1$ motion.

Differential equations in osculating elements for semi-major axis of
a relative orbit $a$ and its eccentricity $e$ are represented as
[10]:
\begin {equation} \label {2.3}
\begin {array} {l}
\displaystyle\frac {da} {dt} = \frac {2a ^ 2} {\sqrt {GMa (1-e ^
2)}} [e\sin v S +
( 1 + e\cos v) T], \\
\displaystyle\frac {de} {dt} = \sqrt {\frac {a (1-e ^ 2)} {GM}}
\left [\sin v S + \left (\cos v + \frac {\cos v + e} {1 + e\cos v}
\right) T\right].
\end {array}
\end {equation}

Perturbing accelerations $S$ and $T$ are happened to be known after
integration of equations (\ref {1.1}):
\begin {equation} \label {2.4a}
\begin {array} {l}
S = \displaystyle \dot {m} \left (\frac {W _ {2x}} {m} + \frac {V _
{c}} {1-m} \right) + GM\tau\frac {\dot m} {\dot v _ c} \, \left
(\frac {x _ 3 +
pm-p} {r _ {23} ^ 3} -\frac {x _ 3 + pm} {r _ {13} ^ 3} \right), \\
[12pt] T = \displaystyle \dot {m} \frac {W _ {2y}} {m} + GM\tau\frac
{\dot m} {\dot v _ c} \, Y _ 3r _ 3, \quad r _ 3 = \frac {1}
{ r _ {23} ^ 3} -\frac {1} {r _ {31} ^ 3}. \\
\end {array}
\end {equation}

Choosing as the independent variable the true anomaly $v$ of stars,
the equations (\ref {2.3}) can be rewritten as follows
\begin {equation} \label {2.5}
\begin {array} {l}
\displaystyle \frac {da} {dv} = \frac {2a ^ 2} {\dot v \sqrt {GMa
(1-e ^ 2)}} [e\sin v S + (1 + e\cos v) T], \\ [14pt] \displaystyle
\frac {de} {dv} = \frac {1} {\dot v} \sqrt {\frac {a (1-e ^ 2)}
{GM}} \left [\sin v S + \left (\cos v +
\frac {\cos v + e} {1 + e\cos v} \right) T\right], \\
\end {array}
\end {equation}
where
$$
\displaystyle \dot v = \sqrt {\frac {GM} {p ^ 3}} (1 + e\cos v) ^ 2
+ \sqrt {\frac {p} {GM}} \left [\frac {\cos v} {e} S -\frac {\sin V}
{e} T\left (1 + \frac {1} {1 + e\cos v} \right) \right].
$$

After averaging on the true anomaly $v$, equations of stars motion
result in the form
\begin {equation} \label {2.6}
\begin {array} {l}
\displaystyle \frac {d\tilde {a}} {d\tilde {v}} = \frac {1} {2\pi}
\int \limits _ 0 ^ {2\pi} \frac {2a ^ 2} {\dot v\sqrt {GMa (1-e ^
2)}} [e\sin v \, S + (1 + e\cos v)
T] dv, \\
[14pt] \displaystyle \frac {d\tilde {e}} {d\tilde {v}} = \frac {1}
{2\pi} \int \limits _ 0 ^ {2\pi} \frac {1} {\dot v} \sqrt {\frac {a
(1-e ^ 2)} {GM}} \left [\sin v \, S +
\left (\cos v + \frac {\cos v + e} {1 + e\cos v} \right) T\right] dv. \\
\end {array}
\end {equation}

Besides if instead of $v$ we choose the new independent variable $m$
with the help of the formula $ \displaystyle \frac {d} {d \tilde v}
= \frac {\dot m} {\dot v _ c} \frac {d} {dm} $, then equations of
motion for averaged elements $\tilde a$ and $\tilde e$ will happened
to be as
\begin {equation} \label {2.7}
\begin {array} {l}
\displaystyle \frac {d\tilde {a}} {dm} = \frac {\dot v _ c}
{2\pi\dot {m}} \int\limits _ 0 ^ {2\pi} \frac {2a ^ 2} {\dot v \sqrt
{GMa (1-e ^ 2)}} [e\sin v S + (1 + e\cos v) T] dv, \\ [14pt]
\displaystyle \frac {d\tilde {e}} {dm} = \frac {\dot v _ c}
{2\pi\dot {m}} \int\limits _ 0 ^ {2\pi} \frac {1} {\dot v} \sqrt
{\frac {a (1-e ^ 2)} {GM}} \left [\sin v S + \left (\cos v + \frac
{\cos v + e} {1 + e\cos v} \right) T\right] dv.
\end {array}
\end {equation}

After calculation of definite integrals in right-hand sites of
equations (\ref {2.7}) we shall receive the final form of
differential equations for determination of secular perturbations of
semi-major axis and eccentricity of the stars relative orbit:
\begin {equation} \label {2.8}
\begin {array} {l}
\displaystyle \frac {d\tilde a} {dm} = \displaystyle \frac {2 +
\tilde e ^ 2} {(1 - \tilde e ^ 2) ^ 2} \, \frac {\widetilde {W} _
{2y}} {m} + 2\tilde a ^ 3\sqrt {1 - \tilde e ^ 2} \, \tau y _ 3r _
3, \\ [10pt] \displaystyle \frac {d\tilde e} {dm} = \displaystyle
\frac {2 + \tilde e ^ 2} {2\tilde a (1 - \tilde e ^ 2)} \, \frac
{\tilde e} {1 + \sqrt {1 - \tilde e ^ 2}} \, \frac { \widetilde {W}
_ {2y}} {m} - \frac {3} {2} \tilde e \,\tilde a ^ 2\sqrt {1 - \tilde
e ^ 2} \, \tau y _ 3r _ 3,
\end {array}
\end {equation}
where the designation $ \widetilde {W} _ {2y} = y ' (v _ 0 + \tau) $
is accepted.

If to put $\widetilde{W}_{2y} = 0$, i.e. not to take into account
jet forces, the equations (\ref {2.8}) are simplified and accept the
form
\begin {equation} \label {2.9}
\begin {array} {l}
\displaystyle \frac {d\tilde a} {dm} = 2\tilde a ^ 3\sqrt {1 -
\tilde e ^ 2} \, \tau y _ 3r _ 3, \\ [10pt] \displaystyle \frac
{d\tilde e} {dm} = \frac {3} {2} \tilde e \,\tilde a ^ 2\sqrt {1 -
\tilde e ^ 2} \, \tau y _ 3r _ 3,
\end {array}
\end {equation}
which admin the existence of such first integral
\begin {equation} \label {2.10}
\tilde a^3 \,\tilde e^4 = \mathrm{const} = \tilde a_0^3 \,\tilde
e_0^4
\end {equation}
with initial conditions $\tilde a_0 = \tilde a (m_0)$ and $\tilde
e_0 = \tilde e (m_0)$.

Further for simplicity the overlined index "tilde"\, above values
$a$ and $e$ will be omitted.

\begin {center}
NUMERICAL RESULTS
\end {center}

The relative orbit of stars was determined by means of numerical
integration of differential equations (\ref {2.8}) and (\ref {2.9}).
The integration was carried out on an interval by independent
variable $m$ from 0.1 up to 0.8.

In fig. 1 are given dependencies of the semi-major axis $ a $ and
eccentricity $ e $ of relative orbit of a accreted-star from value $
q = m / (1 - m) $ for various values of initial radius stars $P_0$
at initial value of eccentricity $ e _ 0 = 0.1 $. From this figure
it is visible, that the change of an orbit strongly depends on
$P_0$. At small values $P_0$ the flowing jet of substance gets on
the surface of a star $ S _ 2 $ with the large velocity and also
creates enough the large jet force accelerating motion of a star,
while for large $ P _ 0 $, on the contrary, the jet force is
directed in the opposite side and consequently drags the motion of
the star.

The influence of initial value of eccentricity $ e _ 0 $ on elements
of star $ S _ 2 $ is represented in fig. 2 for fixed value $ P/a =
0.075 $. As it is seen stars motions strongly depend on the initial
value of eccentricity $e_0$. Under small $e_0$ and $P_0$ secular
changes of the semi-major axes can reach significant values.

In fig. 3 on the plane $ (q, P/a) $ the curves are represented
describing a close binary system. The curve 1 determines border $ P
_ {max} /a = 0.49 \sqrt [3 \;] {m} $, after which the binary forms
contact system, when the sizes of accreted-star reach the singular
Euler point. The solid curves 2-5 for various $ e _ 0 $ determine
border for $ P _ {min} /a $ below of which at close binary system an
accreted disk is formed. Between curves $ P _ {max} /a $ and $ P _
{min} /a $ the half-divided phase of close binary system settles
down. On dot and dash curves 6-9 for various $ e _ 0 $ the first
derivative of the semi-major axis turns into zero. Above these
curves a close binary system extends, and downwords it is
compressed. The model of Paczynski-Huang on this figure is
represented by a vertical straight line $ q = 1 $. According to the
formula ( \ref {V1}) at $ q < 1 $ a binary is compressed, and at $
q> 1 $ it extends, that sharply differs from results which has been
received by means of numerical integration.

In fig. 4 on the same plane $ (q, P/a) $ curves of change are
represented for value $ P/a $ depending on $ q $ for various values
$ P _ 0 $. The curve 1 crosses the border of contact binary system
formation. The further account for this curve is not carried out.
The curves 6 and 7 cross a border of accreted disk formation. If one
considers, that the accreted disk is formed instantly, it is
possible to continue these curves further, using the differential
equations (\ref {2.9}), i.e. believing, that any jet stream does not
create jet force on a star $ S _ 2 $, giving back energy on rotary
motion for particles of accreted disk. Continuations of curves 6-7
on the figure is shown by dashed lines.

In fig. 5 the comparison of change of the semi-major axis and
eccentricity of an orbit of stars is carried out at presence of
accreted disk and at its absence for various initial values of
eccentricity. As it is visible from figure changes $ a $ and $ e $
at presence of accreted disk are not rather great, that is explained
by absence of a jet force and small forces of attraction on stars by
the stream. Therefore for, approximate estimations of evolution of
close stars orbit with presence of accreted disk, taking into
account small sizes of a jet and, hence, small right hand sides of
equations (\ref {2.9}), it is possible to consider, that
\begin {equation} \label {}
a = \mathrm {const}, \quad e = \mathrm {const},
\end {equation}
i.e. instead of Paczynski-Huang model one can use that of
Gylden-Mestschersky.

Dependencies between the semi-major axis $a$ end eccentricity $e$
along trajectories of stars motion are represented in fig. 6. For
the accreted disk these dependencies turn out from the first
integral (\ref {2.10}).

It is difficult to carry out comparison of the present work results
with those of other authors, so as jet forces, created by a stream,
actually nobody takes into account (see, for example, [11]), but in
the present work the jet forces play the basic role.

It is doubtless, that all received results require a further
revision and their conformity to observations.

\begin {center}
CONCLUSION
\end {center}

During the last half-century for definition a relative orbit of
close binary stars the incorrect Paczynski-Huang model was used,
that was marked in the paper [8]. However up to this day in works,
dedicated to close binary systems this model continues to be used.
Therefore in the present work once again, but by means of other
methods, the inaccuracy of this model has been shown.

For determination of relative motion of stars in close binary system
in the present work the numerical integration of equations of motion
is used with taking into account of jet forces and forces of
attraction of stars by the flowing jet. The above calculations of
elliptic orbits of close binary stars show that the eccentricity of
orbits can change strongly enough. The influence of jet force on
orbital evolution of stars can be various. If the accepting star has
smaller mass and little sizes, and the accreted disk is absent, then
the eccentricity is increased strongly, reaching values 0.5-0.8.
Probably just the such influence of jet force explains rather large
number of close double stars with great orbital eccentricity. If the
accepting star has the large mass and the great size, the jet force
creates the braking effect, and the eccentricity decreases down to
zero.

It is shown that for approximate determination of orbital evolution
of close binary system with a generated accreted disk instead of the
Paczynski-Huang model it is necessary to use the model of
Gylden-Mestschersky, according to which all Keplerien elements
remain constant, in particular, $ a = \mathrm {const} $ and $ e =
\mathrm {const} $.

It is doubtless, that obtained results can be specified, taking into
account other perturbation factors and making new assumptions based
on observations.

The present work is maintained by the Russian fund of fundamental
researches (grant 08-02-00398).

\newpage
\begin {center}
REFERENCES\end {center}

 1. A.Kruszevski, Adv. Astron. And Astrophys. {\bf4}, 233 (1966).

 2. J.Hadjidemetriou, Astrophys. And Space Sci. {\bf3}, 330 (1969).

 3. S.L.Piotrovski, Astron.J. (russian), {\bf44}, 241 (1967).

 4. S.S.Huang, Ann. Rev. Of Astron. And Astroph. {\bf 4}, 35 (1966).

 5. B.Paczynski, Acta Astron. {\bf 16}, 231 (1966).

 6. I.W.Mestschersky, Astron. Nachr. {\bf159}, 229 (1902).

 7. T.Levy-Civita, Rendiconti della reale Acad. Lincei. {\bf 8}, ¹
12, 621 (1928).

 8. L.G.Lukyanov, Letter in Astron. (russian), {\bf 31}, ¹ 8, 628
(2005).

 9. L.G.Lukyanov, Astron.J. (russian), {\bf 85}, 755 (2008).

10. G.N.Douboshin,  The celestial mechanics. The basic problems and
methods (Moscow, 1968).

11. L.G.Lukyanov, Astron.J. (russian), {\bf 82}, 1137 (2005).

12. J.F.Sepinsky, B.Willems, V.Kalogera, F.A.Rasio, The Astrophys.
J., {\bf 667} 1170 - 1184 (2007).

\newpage
\begin{figure}[]
\subfigure[]{\includegraphics[width=0.47\textwidth]{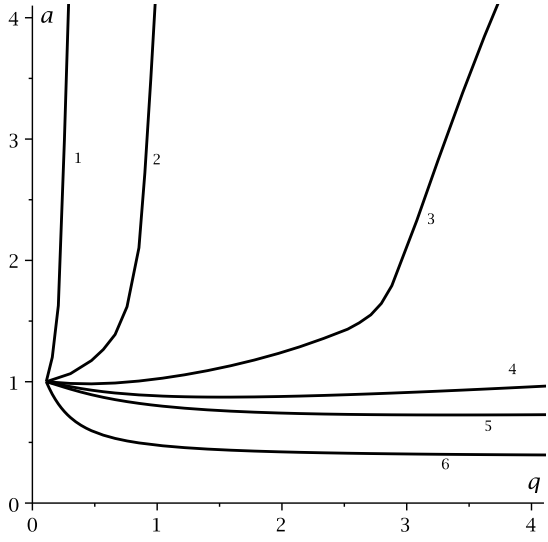}}
\hfill
\subfigure[]{\includegraphics[width=0.47\textwidth]{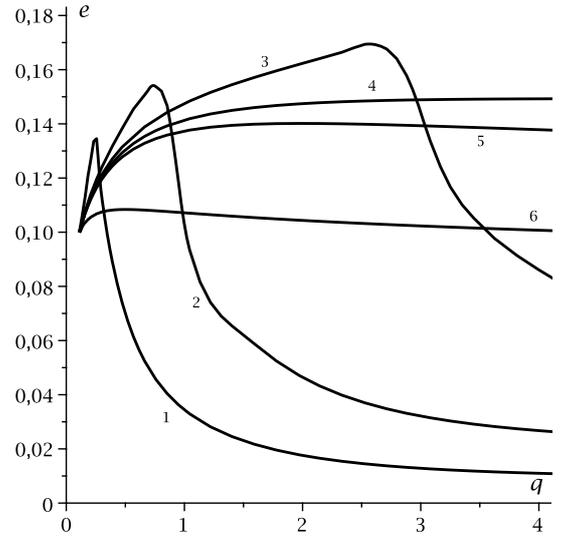}}
\caption{The diagrams of functions: (a) --- $ a (q) $ and (b)
--- $ e (q) $ at $ a _ 0 = 1 $ and $ e _ 0 = 0.1 $
for various $ P _ 0 $. Notations: 1 --- $ P _ 0 = 0.06 $, 2 --- $ P
_ 0 = 0.073 $, 3 --- $ P _ 0 = 0.075 $, 4 --- $ P _ 0 = 0.0758 $, 5
--- $ P _ 0 = 0.0765 $, 6 --- $ P _ 0 = 0.1. $} \label{Fig1}
\end{figure}

\begin{figure}[]
\subfigure[]{\includegraphics[width=0.47\textwidth]{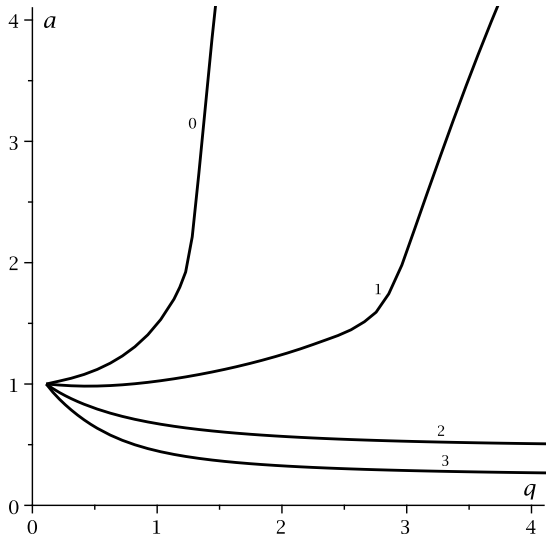}}
\hfill
\subfigure[]{\includegraphics[width=0.47\textwidth]{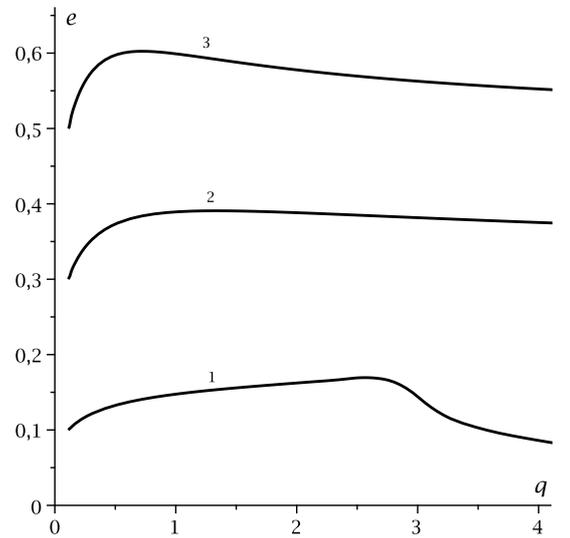}}
\caption{The diagrams of functions: (a) --- $ a (q) $ and (b)
--- $ e (q) $ at $ a _ 0 = 1 $ and $ P _ 0 = 0.075 $ for various $ e _ 0 $.
Notations: 0 --- $ e _ 0 = 0 $, 1 --- $ e _ 0 = 0.1 $, 2 --- $ e _ 0
= 0.3 $, 3 --- $ E _ 0 = 0.5 $.} \label{Fig2}
\end{figure}

\begin{figure}[p]
\includegraphics[width=0.37\textheight]{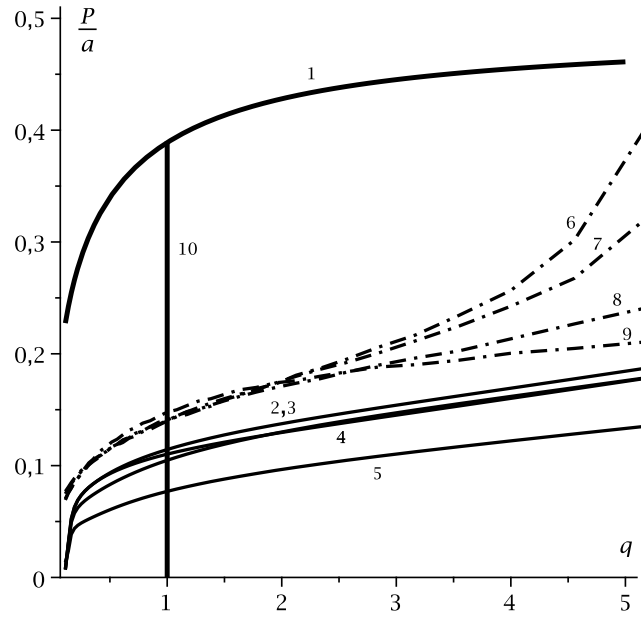} \hfill
\centering\caption{Characteristic curves on a plane $
\displaystyle\left (q, \frac {P} {a} \right) $. Notations: 1 --- the
border of formation of the contact system, solid curves 2, 3, 4, 5
--- the border of accreted disk formations for
values $ e _ 0 = 0 $, $ e _ 0 = 0.1 $, $e_0 = 0.3 $, $ e _ 0 = 0.5 $
accordingly, doted and dashed curves 6, 7, 8, 9
--- the border of change of a sign of derivative $a'$ for the same
values $ e _ 0 $ accordingly, vertical direct line 10 --- the border
of change sign by derivative $ a ' $ under the Paczynski-Huang
formula.} \label{Fig3}
\end{figure}

\begin{figure}[p]
\includegraphics[width=0.37\textheight]{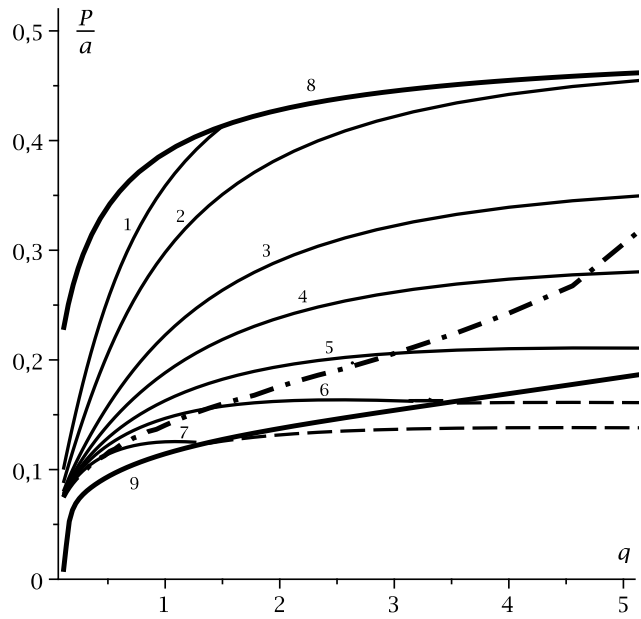} \hfill
\centering
\caption{Dependence of the ratio $ \displaystyle\frac {P}
{a} $ from $ q $ for different trajectories of motion of stars at $
a _ 0 = 1 $ and $ e _ 0 = 0.1 $. Notations: 1 --- $ P _ 0 = 0.1 $, 2
--- $ P _ 0 = 0.0878 $, 3 --- $ P _ 0 = 0.08 $, 4 --- $ P _ 0 =
0.0778 $, 5
--- $ P _ 0 = 0.0765 $, 6 --- $ P _ 0 = 0.0758 $, 7
--- $ P _ 0 = 0.075 $, 8 ---  the border of contact system, 9
--- the border of accreted disk. Dashed curves --- continuation
the appropriate curves after formation of accreted disk, doted end
dashed curve
--- the border of change of sign of derivative $ a ' $.} \label{Fig4}
\end{figure}

\begin{figure}[p]
\subfigure[]{\includegraphics[width=0.47\textwidth]{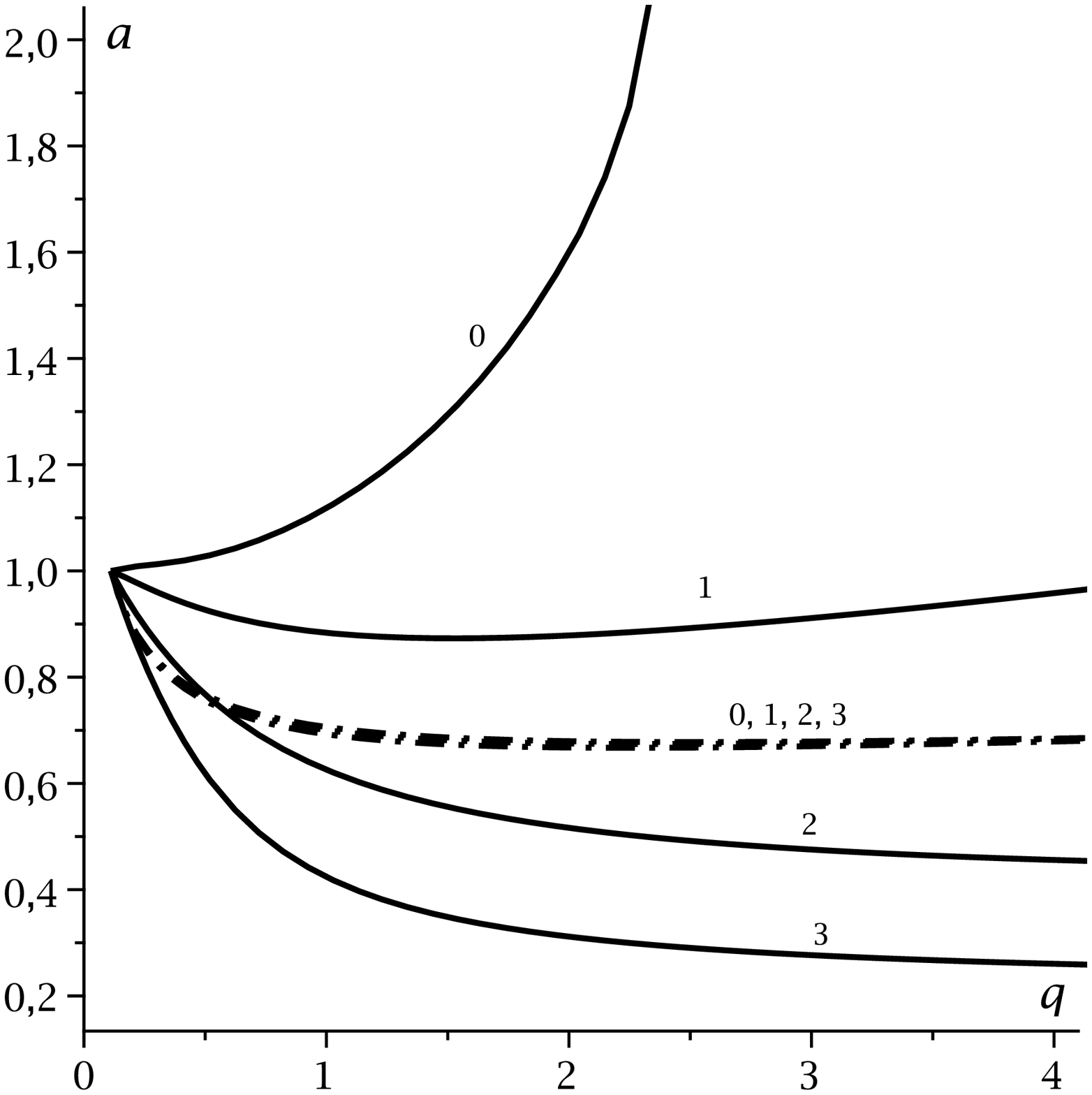}}
\hfill
\subfigure[]{\includegraphics[width=0.47\textwidth]{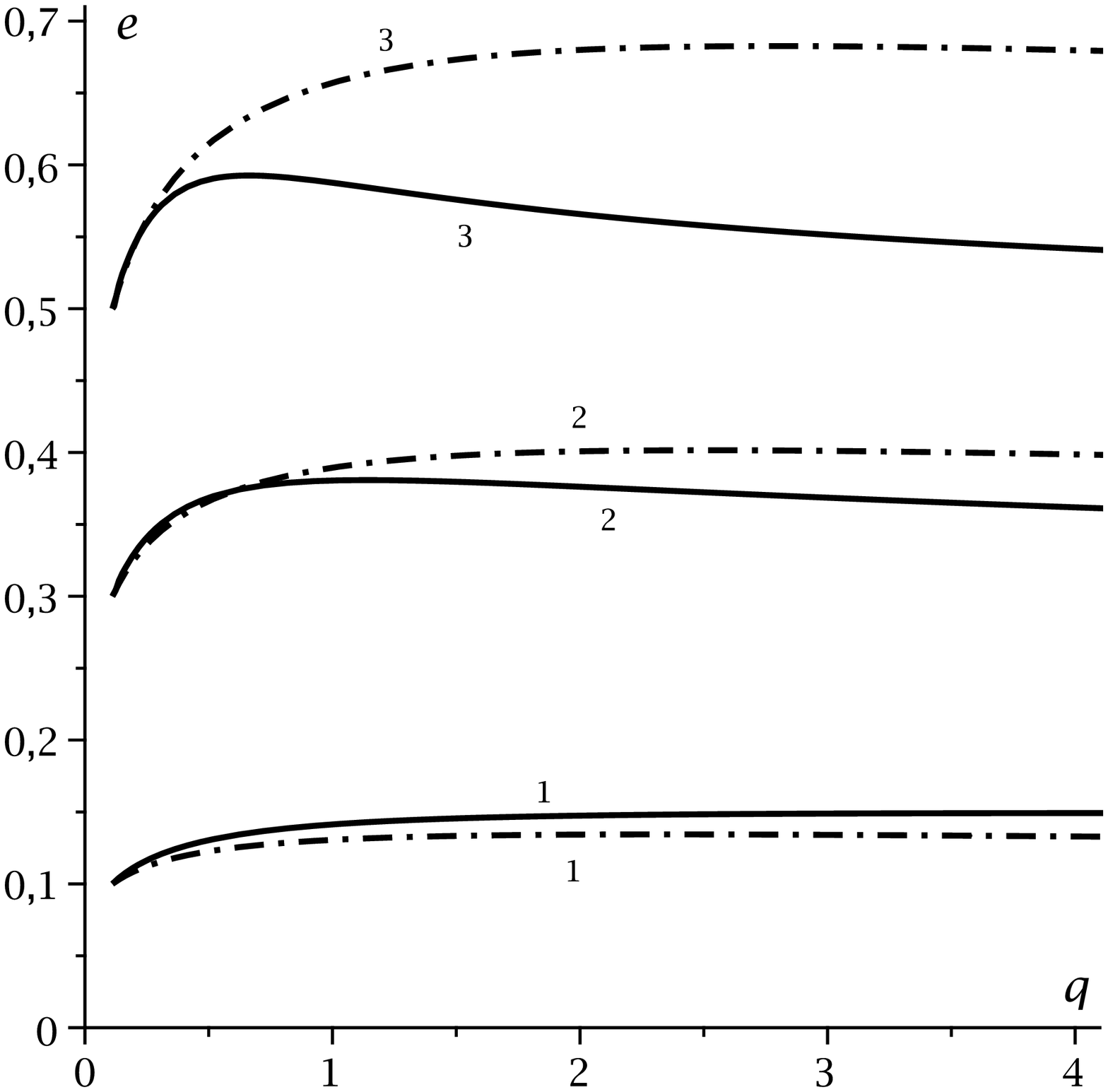}}
\caption{The diagrams of functions: (a) --- $ a (q) $ and (b) --- $
e (q) $ for $ a _ 0 = 1 $ and $ P _ 0 = 0.0758 $. Notations: 0 --- $
e _ 0 = 0 $, 1 --- $ e _ 0 = 0.1 $, 2
--- $ e _ 0 = 0.3 $, 3 --- $ e _ 0 = 0.5 $. Solid curves
correspond to absence of accreted disk, and dot end dash curve
--- to its presence.}
\label{Fig5}
\end{figure}

\begin{figure}[p]
\subfigure[Without of accreted disk at $ a _ 0 = 1 $ and $ e _ 0 =
0.5 $. Notations: 1 --- $ P _ 0 = 0.0758 $, 2 --- $ P _ 0 = 0.075 $,
3 --- $ P _ 0 = 0.073
$.]{\includegraphics[width=0.47\textwidth]{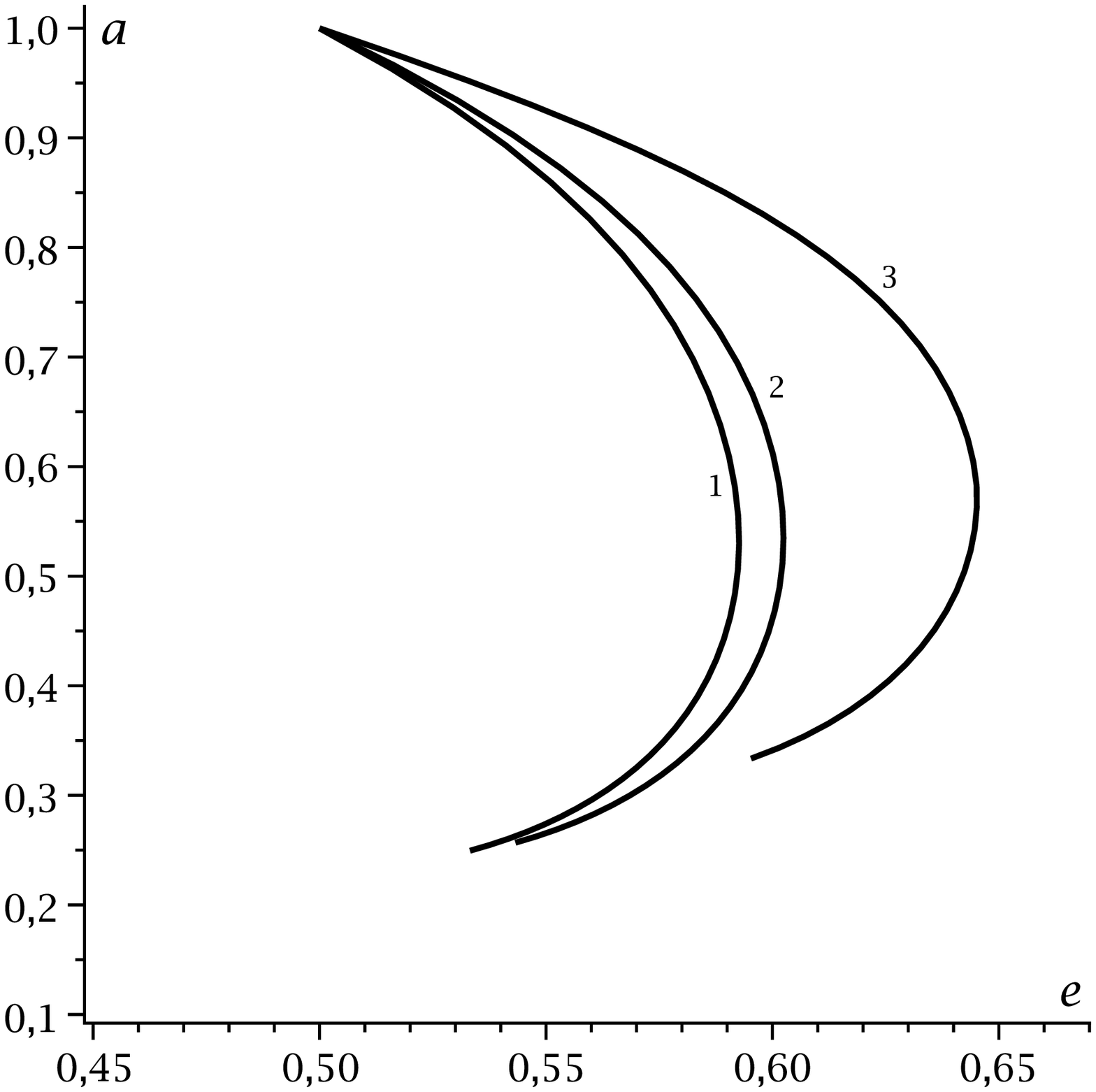}} \hfill
\subfigure[With accreted disk. Notations: 1
--- $ e _ 0 = 0.1 $, 2
--- $ e _ 0 = 0.3 $, 3 --- $ e _ 0 = 0.5 $.]
{\includegraphics[width=0.47\textwidth]{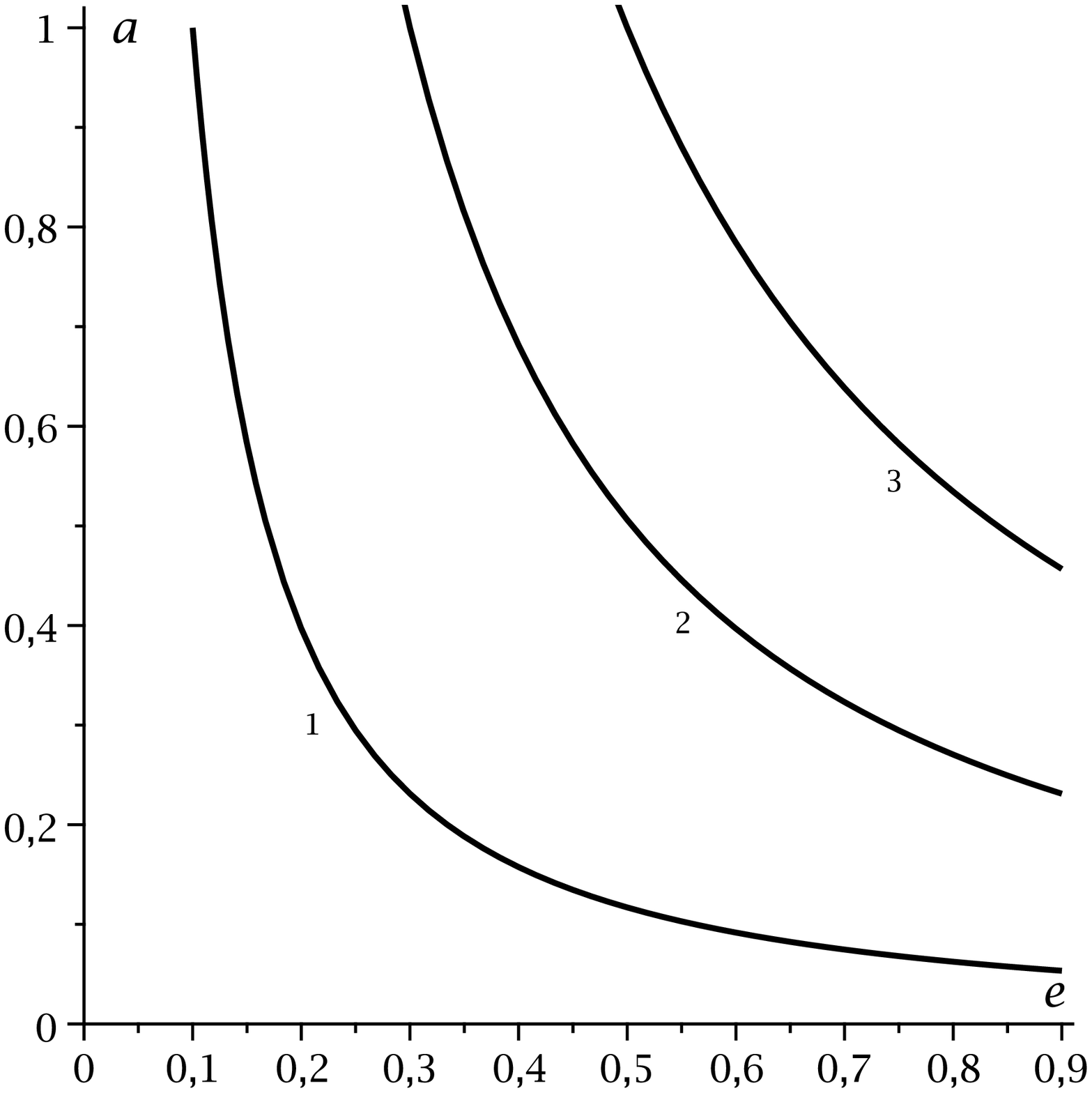}}
\caption{Dependencies between $ a $ and $ e $ along trajectories.}
\label{Fig6}
\end{figure}

\end{document}